# Blockchain in the Quantum World

Arman Rasoodl Faridi, Faraz Masood, Ali Haider Thabet Shamsan, Mohammad Luqman, Monir Yahya Salmony
Department of Computer Science
Aligarh Muslim University
Aligarh, India

*Abstract*—Blockchain is one of the most discussed and highly accepted technologies, primarily due to its application in almost every field where third parties are needed for trust. Blockchain technology relies on distributed consensus for trust, which is accomplished using hash functions and public-key cryptography. Most of the cryptographic algorithms in use today are vulnerable to quantum attacks. In this work, a systematic literature review is done so that it can be repeated, starting with identifying the research questions. Focusing on these research questions, literature is analysed to find the answers to these questions. The survey is completed by answering the research questions and identification of the research gaps. It is found in the literature that 30% of the research solutions are applicable for the data layer, 24% for the application and presentation layer, 23% for the network layer, 16% for the consensus layer and only 1% for hardware and infrastructure layer. We also found that 6% of the solutions are not blockchain-based but present different distributed ledger technology.

*Keywords—Blockchain; quantum computers; distributed ledger technology; security; systematic literature review; quantum attacks*

## I. INTRODUCTION

Quantum computing is one of the latest technologies that has exploded in popularity in recent years. While the foundation of quantum mechanics has been more theoretical than practical for over 100 years, now the time has arrived when practically all firms are delving into it. In the late 1970s and early 1980s, research defining the fundamentals of quantum computing surfaced. Paul Benioff, an Argonne National Labs scientist, wrote a paper in 1979 that showed the theoretical foundation for quantum computing [1] and suggested that a quantum computer might be developed. Numerous businesses claim to be developing quantum computers, such as IBM, which is currently providing its clients with the first solutions in the form of a Quantum Gate Model. Google, Microsoft and many other companies are exploring similar machines.

Satoshi Nakamoto introduced the decentralised transfer and maintenance of digital assets that cannot be duplicated [2]. Distributed ledger technology (DLT) was initially used in finance, but it was subsequently discovered that it could be used whenever we desire to eradicate centralisation or intermediaries. The most widely used DLT is blockchain. There are other types of DLTs like IOTA [3], Hashgraph [4] etc., which are based on Directed Acyclic Graphs. Radix is also a DLT that uses a distributed database to store transactions [5]. Blockchain may be conceived as a sequence of interconnected blocks containing transactions. Every block stores the hash of the previous block, which results in a chain that is very difficult to modify since modifying every transaction necessitates modifying the block, and modifying the block necessitates modifying the entire chain. Blockchain is the foundation of cryptocurrencies such as Bitcoin [2], Ethereum [6], Litecoin, etc.

Quantum computers cannot solve optimisation issues in a substantially scalable manner. In a universal infrastructure, there will be classical computers and quantum computers, with the quantum computer having a significant edge in terms of optimisation. Several quantum algorithms, such as Grover's algorithm [7], Shor's algorithm [8], and others, can solve some problems far quicker than conventional algorithms. Problems that have previously been almost insolvable will now be resolved in a reasonable period. In this regard, advancement in the quantum computing sector has piqued the curiosity of many researchers in both academia and industry.

Blockchain technology started to proliferate because of its nature to provide unbreakable data security, but once practical quantum computers are developed, they cannot provide such security [9]. Smart contracts can be hampered, and the whole technology will go down. The security of the blockchain is built on mathematical challenges that are extremely difficult for even the most powerful conventional computers to solve.

Public key cryptography protects cryptocurrencies. To breach public key encryption, quantum computers might possibly threaten the crypto industry, where some currencies are worth trillion of dollars. Encryption can be bypassed, allowing attackers to mimic legal owners of digital assets. All security assurances will be meaningless if quantum computing gets strong enough. To decrypt data, quantum computers will need thousands of qubits, compared to today's hundreds. Machines will also require persistent qubits that can do calculations for much longer than currently achievable.

NIST (National Institute of Standards and Technology) has already started finding, evaluating, and standardising public-key cryptography algorithms that are quantum-resistant [10]. However, it is necessary for the research community to primarily focus on blockchain technology. A lot of work is going on to create a quantum secure blockchain. To systematically analyse them following research questions are set:

RQ1: What challenges and security issues could occur due to the rise of quantum computers in blockchain technology?

RQ2: What are the various strategies and approaches used by researchers to make blockchain quantum resistant?

To answer these research questions, a systematic literature review has been undertaken. In Section 2, the research method is discussed in detail. Section 3 explores the basics of blockchain and quantum computing and the related challenges and solutions associated with these technologies. The survey results and answers to the research questions are discussed in Section 4, and the work is then concluded in Section 5 with future directions.

## II. RESEARCH METHOD

This research utilised the SLR (Systematic Literature Review) method, as it helps to conduct the secondary research using a well-defined method. This approach gives us a framework to follow in order to discover, analyse, and evaluate relevant literature to find unbiased and reproducible answers to our research questions [11]. The parts of the process include planning, conducting and reporting on the review. Section 1 deals with the planning phase. Reporting is handled in Sections 3 and 4. This section goes through the phases of the review process, which includes:

### A. Research Identification

This preliminary search aims to discover existing systematic reviews and determine the volume of studies that would be appropriate. A single search string is utilised instead of many search strings. Only databases related to the issue and widely accepted in the scholarly community are included. For this study, only IEEE Xplore, Elsevier, ACM Digital library, Springer Nature and Taylor & Francis is used.

The search string is formed to search throughout the metadata using the Boolean operator "AND," and the simple search term is ("Blockchain" AND "Quantum").

### B. Inclusion and Exclusion Criteria

Inclusion and exclusion criteria should be based on the research questions to guarantee that the research questions can be effectively interpreted and that the studies are properly classified. Because the wide usage of blockchain grew in prominence after 2015, we chose all papers published after January 2016. We also limited the results to journal and conference articles, excluding online material, books, and magazines. If duplicate articles or corrections are found in any of these articles, they were removed. Finally, only articles written entirely in the English language are chosen.

### C. Study Selection Process

We found 126 items in IEEE Xplore, 395 in Elsevier, 187 in the ACM digital library, 272 in Springer Nature, and 96 in Taylor & Francis Online using the specified search term and inclusion-exclusion criteria. A three-stage selection technique was implemented to guarantee that only relevant research articles were evaluated. Following the search, the results are extracted using keywords and titles. Following that, the abstracts of the papers were read, and the number of articles was decreased. Only high-quality studies that answered the research questions were picked in the last step, which involved reading whole articles and ranking them based on content. For efficient monitoring and control during the selection process, a separate folder was created for each evaluation stage, along with a new Excel sheet. The research is entirely transparent and traceable as a result of this. Table I shows the step-by-step selection criteria, and Fig. 1 shows the count of publications that were included.

TABLE I. CRITERIA FOR ACCEPTANCE AT EACH STAGE

| Review Stage | Method | Criteria for acceptance |
|---|---|---|
| First | Filter the articles using keywords and titles. | The title or keyword should be related to the research objective. Select the document for the next stage if there is any doubt. |
| Second | Exclude articles based on abstracts | Check if the abstract relates to the research question. In case of doubt, move the paper to the subsequent stage. |
| Third | Articles are excluded based on their entire text and article quality. | Papers that correspond to the research subject and proper experiments or mathematical proof is provided are selected. |

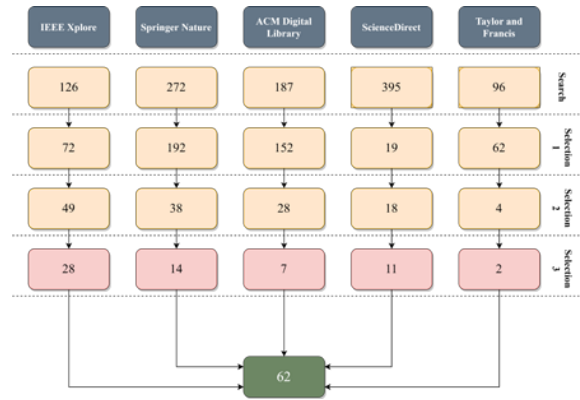

Fig. 1. Number of Papers Selected at each Stage.

### D. Data Extraction

Once the analysis of the selected articles was completed, an excel file was created to record the data extracted from each publication. Table II shows the fields that were taken from each publication.

TABLE II. FIELDS USED FOR DATA EXTRACTION

| S. No | Field | Description |
|---|---|---|
| S1 | Title | The paper's title |
| S2 | Database | Where an article is published |
| S3 | Rating | According to the content |
| S4 | Experiment | Whether or not proper experimentation is carried out |
| S5 | Mathematical Proof | Whether or not mathematical proof is provided |
| S6 | Architecture/Framework/Algorithm | Whether the architecture, framework, or algorithm is given. |
| S7 | Code | Whether source code is given to duplicate the results |
| S8 | Survey | Is it a survey paper |
| S9 | Problem identified | Which type of issue is discussed in the paper |
| S10 | Category of Solution | What type of solution is provided |

*E. Data Synthesis*

According to the research questions, all data taken from the selected publications was synthesised. This makes it simple to understand the challenges and different kinds of solutions provided.

As shown in Tables III and IV, a systematic data analysis assisted in the formalisation of specific categories related to the description of problems and solutions.

TABLE III. CHALLENGES BASED ON LAYERS

| S. No. | Layers | Articles |
|---|---|---|
| 1 | Application and Presentation Layer | [17], [18], [27], [19]–[26] |
| 2 | Consensus Layer | [28]–[36] |
| 3 | Network Layer | [37], [38], [47], [39]–[46] |
| 4 | Data Layer | [48], [49], [58]–[65], [50]–[57] |
| 5 | Hardware and infrastructure layer | [48] |
| 6 | Not based on Layers | [66], [67] |

TABLE IV. SUMMARY OF SOLUTIONS FOUND IN THE LITERATURE

| S.No. | Solution | Article |
|---|---|---|
| 1 | Quantum Properties | [18], [21], [89], [26], [28], [31], [41], [44], [46], [70], [88] |
| 2 | Hash Based Signature | [24], [25], [50], [56], [58], [59], [71] |
| 3 | Code-Based Cryptography | [22] |
| 4 | Lattice Based Cryptography | [20], [23], [53], [54], [57], [62], [63], [90], [38]–[40], [43], [45], [49], [51], [52] |
| 5 | Multivariate Cryptography | [37], [55], [64] |
| 6 | Directed Acyclic Graph | [66], [67] |
| 7 | Quantum Blind Signature | [18], [38], [42], [55], [70] |
| 8 | Quantum Walks | [61] |
| 9 | Hardware And Software Based Blockchain | [48] |
| 10 | Quantum Cloud Computing | [17], [48] |
| 11 | Post-Quantum Threshold Signature | [29] |
| 12 | Quantum Random Oracle Model | [43] |
| 13 | One Way Function | [60][65] |
| 14 | Zero Knowledge Proof | [47][27] |
| 16 | New Consensus | [21], [28]–[30], [32]–[36] |
| 17 | Review | [9], [91]–[94] |

Furthermore, as shown in Fig. 3 and 4, a frequency analysis is performed for the problems and solutions under study.

## III. SLR FINDINGS

We synthesised the data from the selected papers depending on the research questions. The problems are not clearly defined but presented as an overall solution to blockchain problems with quantum computing. In order to categorise them properly, the solutions provided are split based on different layers of blockchain. Every layer has different security requirements, so based on these layers, research articles are grouped. Also, some solutions are working of more than one layer, so these solutions are identified separately for each layer. First, we explain the problems in each layer and then different types of solutions studied in the literature.

*A. Challenges and Issues*

After the analysis, it is decided to represent blockchain in layers as shown in Fig. 2 and then understand the issues according to each layer. Dividing into layers make it easy to understand where research is still required. These layers, along with problems, are explained below:

*1) Hardware and infrastructure layer*: Internet users (peers) can now connect with other peers and share data as distributed systems are becoming more prevalent. This layer is responsible for creating virtual resources such as storage, networks, and servers. Nodes are the essential part of this layer because nodes are hardware devices that connect to the network and help make consensus in the blockchain. Infrastructure security frequently necessitates either limiting or prohibiting access to the node. So, improvement is needed at the infrastructure level to implement quantum blockchain properly.

*2) Data layer*: Data stored in blockchain depends on the type of blockchain-like Hyperledger Fabric [12] that contains channel information, whereas a Bitcoin blockchain needs to store the information about the sender, receiver, and amount. Blockchain network data is added only when consensus is reached among the nodes. Hash functions help in the easy identification of blocks and the detection of any changes made to the blocks. To ensure the confidentiality and integrity of the data stored on the blockchain, transactions are digitally signed. Blockchain uses asymmetric cryptography to secure information about the block, transactions, and transacting parties, among other things.

To sign a transaction, private keys are used, and anyone with the public key is used, and anyone with the public key can verify the signer. Because the encrypted data is also signed, digital signatures ensure data integrity. Every transaction in a block is hashed and organised in the form of a Merkle tree. In the Merkle tree hash of transactions are organised in the form of a binary tree. If any transaction is changed, then the whole Merkle tree is changed, which changes the whole block as the block contains the hash of the Merkle tree.

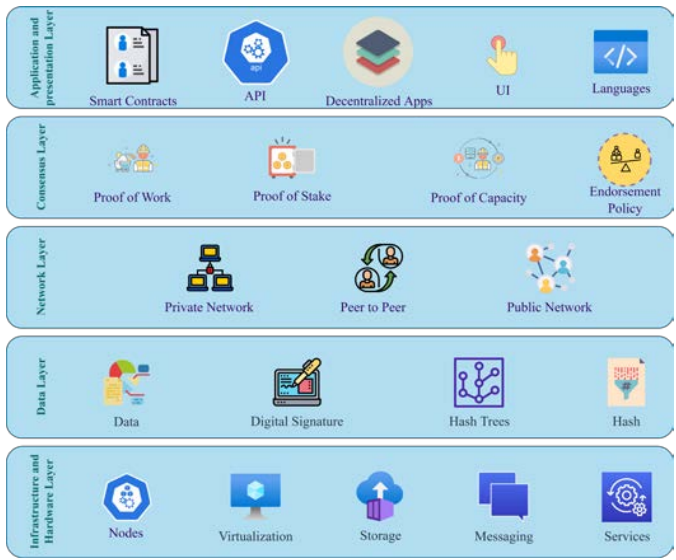

Fig. 2. Different Layers of the Blockchain.

As a result, any manipulation will render the signature invalid. Most blockchain systems depend significantly on a digital signature to improve security. These signatures rely on the difficulty of solving a mathematical problem, such as determining the factors of large integers. The data layer is highly dependent on these algorithms, and once practical quantum computers are developed, breaking these algorithms will be easy. As a result, this layer is too much vulnerable to quantum attacks.

*3) Network layer*: The network layer is in charge of inter-node communication and handles block propagation, transactions, and discovery. It is also called the propagation layer or peer to peer layer. In a peer-to-peer network, nodes share the workload to achieve a common goal in a distributed network. This layer ensures that nodes can discover other nodes in the network to interact, propagate, and synchronise information with other nodes. This layer also handles the propagation of the world state. A node can be a light node or a full node. Light nodes can merely retain the blockchain's header and send transactions. Full nodes are responsible for transaction verification and validation, mining, and consensus rule enforcement. They are in charge of ensuring the network's trustworthiness. So, it is needed that this layer uses quantum network in the future.

The term "Quantum Internet" [13]–[15] refers to the entire system, which comprises both quantum and classical packet switching networks. A traditional network in which hosts and routers can handle quantum information in the network graph structure is known as a quantum network. Between these nodes, there are classical channel for transferring classical data and quantum channel connections for transmitting quantum data.

*4) Consensus layer*: The rules that nodes follow to ensure that transactions are validated within those rules, and that blocks respect those rules is known as consensus. There is a consensus algorithm behind every blockchain as the trusted third party is missing to validate transactions in case of conflict. The consensus layer is the most significant layer for any blockchain. Consensus protocols provide a set of irrefutable agreements between nodes in a distributed peer-to-peer network. Consensus keeps all of the nodes in sync. Consensus is in charge of validating the blocks, ordering them, and guaranteeing that everyone agrees. It is easy to attack this layer with the help of quantum computers. Attackers can search hash collisions, which can subsequently be used to change blocks in a network without impacting the integrity of the blockchain. Also, for mining, it is required to search a nonce, and with quantum computers, it will be very fast. This can enable an attacker to reconstruct the whole blockchain without getting detected by the network.

*5) Application layer*: The application layer includes smart contracts[16], chaincode[12], and decentralised apps (dApps). Smart contracts are digital contracts built on the blockchain that is automatically executed when particular events occur, or any external criterion are met. Chaincode is a collection of related smart contracts used to do a certain purpose. dApps are software applications that run on a blockchain network of computers rather than on a single device. Because they are decentralised, decentralised apps are free of the control and influence of a single authority.

dApps provide several advantages, including user privacy, developer independence and lack of censorship. The application layer comprises two layers: the application layer and the execution layer. The application layer is where end-users interact with the blockchain network, including scripts, APIs, user interfaces, and frameworks.

The execution layer, which includes smart contracts, underlying rules, and chaincode, is a sublayer. This sublayer contains the code and rules that are actually executed. A transaction is propagated from the application to the execution layer, but it is validated and executed by the semantic layer. The execution layer processes transactions and preserves the blockchain's deterministic nature. It receives instructions from the application layer. A smart contract code should not be pulled down or changed after being deployed on a blockchain, but it may be possible with quantum computers. Similarly, instead of making the smart contract more complex within the same technology, it is necessary that from now on, those researchers should start moving towards quantum-resistant smart contracts.

*B. Solutions*

First, the basic concepts of quantum computing and some of the properties are discussed in this section. After that, the explanation of the solutions and methods that are found in the literature are discussed.

Quantum computing focuses on developing computer systems using quantum theory and quantum bits, or qubits. Quantum computers use subatomic particles' ability to exist in many states, i.e., it can be 0 or 1 simultaneously. Algorithms work by manipulating bits with gates, which change their states. The NAND gate is a universal gate, but the NAND gate's behaviour is not reversible because it accepts two inputs and returns outputs that are not unique. In quantum computing, working with reversible gates is typically convenient since every reversible gate may be implemented on a quantum computer. The Toffoli gate is a reversible gate that takes three bits as input, can imitate the NAND gate.

Toffoli's gate converts $(α, β, γ)$ to $(α, β, (γ + α * β) \mod 2)$. The Toffoli gate maps $(α, β, 1)$ to $(α, β, α \text{ NAND } β)$ when $γ = 1$. Quantum computers are classically computationally ubiquitous because they can implement the Toffoli gate, even if the Toffoli gate alone is insufficient to implement any function on quantum states. The electron, which can have a spin pointing up or down, provides a simple physical prototype for this two-state system. These states are usually written as $|0⟩$ and $|1⟩$ in quantum mechanics as a convention. Quantum computers, unlike conventional computers, are not limited to manipulating only these two states. State superpositions, such as $\frac{|0⟩+|1⟩}{\sqrt{2}}$ are also feasible. These two-state systems are known as quantum bits or qubits. Qubit states can alternatively be represented as two-dimensional vectors, for example.

$$|0⟩ = \begin{pmatrix} 1 \\ 0 \end{pmatrix}, \quad |1⟩ = \begin{pmatrix} 0 \\ 1 \end{pmatrix} \quad (1)$$

This is significant because multiplying the corresponding vectors allows gates to be represented mathematically as $2 \times 2$ matrices acting on qubits.

A linear combination of $|0⟩$ and $|1⟩$ with complex coefficients can be used to describe the state $|ψ⟩$ of any given qubit, i.e.

$$|ψ⟩ = p|0⟩ + q|1⟩, \quad p, q \in \mathbb{C}. \quad (2)$$

A classical computer requires two complex numbers to describe an arbitrary quantum state; similarly, modelling n arbitrary quantum states on a classical computer requires $2^n$ complex numbers and so a minimum of $2^n$ Bits. By definition, a quantum computer requires just n qubits to describe $n$ states. Modelling quantum systems on the classical computer will thus take the time that grows exponentially with the number of states $n$, whereas modelling the same system on a quantum computer only requires time that grows linearly with $n$. In other words, the classical computer takes $O(2^n)$ time and the quantum computer takes $O(n)$ time for this example.

The Hadamard gate is important in quantum computing. This gate, represented by $H$, has the following representations in matrix form and state notation:

$$H = \frac{1}{\sqrt{2}} \begin{pmatrix} 1 & 1 \\ 1 & -1 \end{pmatrix} = \frac{1}{\sqrt{2}} \sum_{a,b \in \{0,1\}} (-1)(-1)^{ab} |a⟩⟨b| \quad (3)$$

The Hadamard gate is a crucial component of quantum algorithms like Shor's algorithm, Grover's, and Simons algorithm as the Hadamard gate translates $n$ qubits that are all in the same state to an equal superposition of the $n$ qubits' potential states. Shor's algorithm, on the one hand, argues that, due to quantum mechanics, factorisation may be done in polynomial time rather than the exponential time, which is the basis of many public-key algorithms. Grover's algorithm, on the other hand, can cut the sufficient security strength of algorithms like the AES (Advanced Encryption Standard) in half for a given key length, rendering infrastructures secured by them open to attack [7].

Shor's algorithm is noteworthy because it solves the complex problems of integer factorisation. The best extant algorithm for this problem is known as the generic number field sieve, and it operates in $O(e^{(\log(N)^{1/3} \, poly(\log \log N))})$, where poly is a complex polynomial. Shor's algorithm outperforms $O((\log N)^3)$ in terms of speed.

Shor's approach employs the well-known Euclidean algorithm to compute the greatest common divisor (GCD) and then Simon's algorithm to gain the exponential speedup. Given an integer Z, one can compute $\gcd(f, Z)$ by selecting a random number $f < Z$. The problem is solved if $f$ is a factor of Z; otherwise, $\gcd(f, Z)$ must equal one. Assume M is the order of $f$, i.e., the lowest positive integer is M such that $f^M \mod N = 1$. Then, as long as M is even and $f^{M/2} + 1$ is not a multiple of Z (which is very likely), both $\gcd(f^{\frac{M}{2}} + 1, Z)$ and $\gcd(f^{\frac{M}{2}} - 1, Z)$ are factors of Z.

The problem is solved as long as the M matching to a particular f can be found. Consider the function $g(x) = f^x \mod Z$ to compute M. The task of computing M is thus reduced to period-finding for this function g since:

$$g(x + M) = f^{x+M} \mod Z = f^M f^x \mod Z = f^x \mod Z = g(x) \quad (4)$$

The problem is solved by using Simon's algorithm. Simon's approach solves the period-finding problem, that is, calculating the period M of a function g that satisfies $g(x) = g(x + M)$ for any x. This was a significant subproblem in Shor's method that provided an exponential speedup: Simon's algorithm runs in $O(n)$ on a quantum computer and $O(2^{n/2})$ on a classical computer [68].

With a uniform superposition of states over $n$ qubits, Simon's algorithm computes the function g on the superposition, measures the answer, and applies the Hadamard gate to the $n$ resultant qubit states. If the period M is represented as a $n$-bit vector $\vec{M}$, measuring the state after applying the Hadamard gate returns a vector orthogonal to $\vec{M}$ with a high probability. After $O(n)$ iterations of this process, one receives $n - 1$ orthogonal vectors to $\vec{M}$. Because $\vec{M}$ exists in an n-dimensional vector space, this is enough to determine the period M.

Grover's algorithm[69] is intended to tackle the problem of unstructured search. This problem can be described formally: given a function that transforms N-digit binary values to either 0 or 1, find x. Grover's algorithm is relatively straightforward to implement. To begin, use the Hadamard gate on a set of $n$ qubits to generate a uniform superposition of states, where $N = 2^n$. Following that, a gate is built that rotates the uniform superposition towards the state $|a⟩$ corresponding to a. With a high probability, measuring the state after $O(\sqrt{N})$ applications of this gate will yield $|a⟩$. This is an improvement over the $O(N)$ steps a random classical algorithm would take to find a best-case situation.

*1) Quantum properties*: A quantum state is a mathematical object that offers a probability distribution for each potential measurement of a system's outcomes. When we combine quantum states, we get another quantum state. Pure quantum states cannot be expressed as a mixture of other states, whereas mixed quantum states can be described as a combination of other states. Quantum computing performs computations by utilising the collective characteristics of quantum states, such as superposition, collapse, and entanglement.

A superposition of quantum states may be thought of as a linear combination of many quantum states, resulting in the development of a new valid quantum state. The basic states are $|0\rangle$ and $|1\rangle$. All the Qubits are superposition on these basic states. Quantum superposition differs substantially from classical wave superposition. A superposition of $2^m$ states, ranging from $|0000…0\rangle$ to $|1111…1\rangle$ will exist for a quantum computer with $m$ qubits. The probability of a quantum state $|\psi\rangle$ is $|A_v|^2$ for any set of values v with probability amplitudes $A_k \in \mathbb{C}^5$ in such a way that $|\psi\rangle := \sum_v A_v |\psi_v\rangle$ for the measurement of $|\psi\rangle$ resulting in $\psi_v$. Authors in [28] discussed the new consensus algorithm using quantum entanglement.

When one particle's quantum state cannot be characterised independently of the other particle's quantum state, they are said to be entangled. Even if the individual components are not in a defined state, the system's quantum state as a whole may be characterised. When two qubits become entangled, a one-of-a-kind relationship is established. The entanglement will be demonstrated by measurements, which may produce a value of 0 or 1 for individual qubits where the measurement of both the qubits will be the same. Even if the particles are separated by a significant distance, this is always true. For a quantum state $|\psi\rangle$ with $|\psi\rangle := \sum_v A_v |\psi_k^X, \psi_k^Y\rangle$, then on measurement of $|\psi\rangle$ then probability X sees $\psi_k^X$ and Y sees $\psi_k^Y$ is equal to 1.

While interacting with the outside environment, any wave function is reduced to a single eigenstate from the superposition of many eigenstates, and then it is called wave function collapse. In this case, the probability is 1 for all measurements of quantum state $|\psi\rangle$ resulting in $\psi_v$ where $|\psi\rangle := \sum_v A_v |\psi_v\rangle$, for some v.

A quantum channel can transfer both quantum and classical information. Quantum channels are trace-preserving mappings between spaces of fully positive operators. In other words, a quantum channel is just a quantum operation considered as a pipeline meant to transmit quantum information rather than simply the reduced dynamics of a system. Some solutions, as discussed in [18], [65] are based on quantum channels.

The idea of quantum key distribution (QKD) was initially presented in the 1970s, but it was not fully realised until the 1980s. QKD allows to sharing and distribute secret keys for cryptographic protocols. The essential thing is to keep them private, just between the communicating parties. Quantum superpositions or quantum entanglement and conveying information in quantum states may be used to develop a communication system that detects eavesdropping. If the extent of eavesdropping is less than a certain threshold, only then a secure key can be generated otherwise, the communication is terminated. This is the general concept of Quantum cryptography that is why it is added as a property. Authors in [18], [19], [21], [26], [31], [41], [46], [70] discussed the usage of QKD for quantum blockchain.

*2) Hash-based signature*: The hash-based signature is used to utilise the cryptographic safe hash function properties. These properties include pre-image resistance, one-wayness and collision resistance. Hash-based signature systems rely entirely on the underlying safe cryptographic hash function, limiting the attack surface and cryptanalysis possibilities. By removing the need for several security components, hash-based signature systems substantially minimise implementation complexity. Any hash function that meets the security criteria of cryptographic hash functions can be employed to build hash-based signature algorithms. Because of this inherent flexibility, several underlying hash functions may be used to meet the required performance requirements based on the application-specific environment. Any difficult-to-invert function may be converted into a secure public-key signature system using hash-based cryptography. As a result, this might be a solution for post-quantum blockchains as discussed in [24], [32], [50], [71].

*3) Code-based cryptography*: All cryptosystems, symmetric or asymmetric, whose security is based, in part or entirely, on the difficulties of decoding a linear error-correcting code, perhaps chosen with some particular structure or in a particular family (for instance, quasi-cyclic codes, or Goppa codes) is code-based cryptography [72]. The ciphertext is a codeword with flaws that can only be corrected by the owner's private key (the Goppa code). Grover's algorithm does not significantly outperform earlier code-based cryptosystem attacks in terms of speed.

*4) Lattice-based cryptography*: A lattice is a collection of points having a periodic structure in $n$-dimensional space. Given $n$-linearly independent vectors $v_1, v_2, v_3, ……, v_n \in \mathbb{R}^m$ the set of vectors created by them is the lattice $\mathcal{L}$

$$\mathcal{L}(v_1, v_2, v_3, ……, v_n) = \{\sum x_i v_i \mid x_i \in \mathbb{Z}\} \qquad (5)$$

A basis of the lattice is made up of the vectors $v_1, v_2, v_3, ……, v_n$.

Because of its strong security proofs based on worst-case hardness, reasonably efficient implementations, and considerable simplicity, lattice-based cryptography [73] appears to promise post-quantum cryptography. In two ways, the worst-case security guarantee is critical. It helps us determine the cryptosystem's concrete parameters by ensuring that the cryptographic framework is free of fundamental flaws.

*5) Multivariate cryptography*: A set of (usually) quadratic polynomials over a finite field is a public map for a multivariate public-key cryptosystem (MPKC) [74]. In general, finding a solution to such structures is an NP-complete/-hard problem [75]. One of the intriguing instances is Patarin's Secret Fields [76], which generalises a suggestion by Matsumoto and Imai [77]. The NP-hardness of solving nonlinear equations over a finite field underpins its fundamental security assumption. This is one of the most influential families of PKCs (public-key cryptography), as it can withstand even the most powerful quantum computers in the future. The MQPKC, unlike many other forms of PKC, cannot be solved quickly using Shor's algorithm with a conventional computer because it does not rely on any of the difficulties that Shor's algorithms can resolve.

*6) Directed acyclic graph*: A distributed ledger technology, a DAG [66], is an alternative to regular blockchain that seeks to solve blockchain technology's speed, scalability, and cost concerns. DAG is also a system that uses a digital ledger to keep track of transactions. DAG (Directed Acyclic Graph) is a more expressive outline than an entirely linear model. A DAG is a data or information structure that may be used to show a variety of difficulties. It is a topologically ordered acyclic graph. The node follows a specific sequence for each directed edge. Every DAG begins with a node with no parents and ends without children. There are no cyclic graphs on this page. A DAG is made up of nodes and arrows that connect them. By allowing many chains to exist on the system simultaneously, DAG can solve the single-chain problem of blockchain. IOTA is a DAG currency that is quite well-known. DAG Tangle is what they call it. It eliminates the need for miners in the verification process entirely. The white paper published by IOTA claims that Tangle is quantum-proof [3].

*7) Quantum blind signature*: A blind signature is a digital signature that blinds the message before it is signed. As a result, the message will go undetected by the signer. After that, the signed message will be unblinded. It functions as a standard digital signature and can be publicly verified. Blind signatures that can survive quantum attacks are referred to as "post-quantum blind signatures." Blind signatures have been widely used in the applications like the creation of e-cash and voting agreements. As a result, new quantum blind signature technologies will be necessary for the future. This solution works with other solutions like lattice-based or multivariate cryptography in order to provide quantum-resistant blockchain [18], [19], [39], [42], [54], [55].

*8) Quantum walks*: A random walk is a random process in mathematical space that defines a path consisting of a series of random steps, as defined by Pearson in 1905 [78]. Random walks are essential in solving practical issues since they can be used to evaluate and mimic the unpredictability of items and determine the correlation between them. Quantum walks were introduced in 1993 [79]. The polar opposite of traditional random walks is quantum walks. Quantum walks differ from regular random walks in that they do not converge to any limiting distributions and are much faster because of Quantum interference[79]–[81]. Quantum walks can outperform any traditional algorithm by order of magnitude. The two types of quantum walk-based algorithms are continuous time-based and discrete time-based algorithms [82].

*9) Hardware and software based blockchain*: As shown, blockchain implementation may be implemented into many different technology stack layers. So, hardware-based security is also essential. It may involve hardware-based secure key storage or hardware replacement for quantum channels. Hardware-based key storage is already being developed as cold wallets, but it must also be quantum secure. The authors in [48] develop a quantum computing device as a multi-input multi-output quantum channel.

*10) Quantum cloud computing*: In a cloud computing environment, a cloud quantum computer is a computer that can be accessed over the internet. Users may now make use of a variety of cloud quantum computing services to solve complicated issues that demand a lot of computational power. The design and performance of different cloud quantum computing systems vary. Solutions discussed in [17], [48] used quantum cloud computing.

*11) Post-quantum threshold signature*: Threshold signature [83] is a unique digital signature that can be used to identify a group of users. It is generated by an authorised subset of the private keys. The public keys are already generated with these private keys. It is very easy to verify these signatures as only a single public key and a single signature is enough. If at least n users out of m users efficiently sign the message, then the system is known as (n, m) threshold. The solution discussed in [29] is based on solving quadratic equations in a finite field, an NP-hard problem. This system is a threshold signature system and is considered safe even after developing a powerful quantum computer.

*12) Quantum random oracle model*: In a random oracle[84], anyone may give it an input and output of fixed length. If someone has already requested the input, the oracle will provide the identical result. If the oracle receives an input that it has not seen before, it generates a random output. To make the whole system secure, it is needed to replace all the hash functions used in the system with random oracles. Traditional oracle models can be easily attacked by using quantum superposition. This may result in the failure of many classical security proofs and must be rewritten. Quantum random oracle along with lattice-based solutions are discussed in [43].

*13) One way function*: A one-way function is easy to compute on all inputs but complex to invert given the image of a random input. In many cryptographic systems, one-way functions have proven useful primitive. Extensive work on one-way quantum functions has also been done in the post-quantum period. These one-way functions accept outputs of the quantum states by taking classical bit strings as input. Many information-theoretically secured digital signature techniques rely on the one-wayness characteristic of these functions [85], [86] have been proposed. To authenticate both classical bit strings and quantum states, these one-way functions should be both quantum-classical and classical-quantum in design. As a result, [60] developed quantum money systems based exclusively on the security of one-way functions that are resistant to quantum attacks.

*14)Zero-knowledge proof*: Zero-Knowledge Proofs (ZKPs) enable data to be validated without disclosing the data itself. As a result, they have the potential to transform the way data is gathered, used, and transacted. Each transaction is assigned a 'verifier' and a 'prover'. In a ZKPs transaction, the prover tries to prove something to the verifier without revealing anything about it. The authors of [47] suggest employing two indistinguishable hash functions combined with ZKPs protocols to ensure security against quantum attacks.

## IV. RESULT AND DISCUSSION

Blockchain is an up-and-coming technology, and it is assumed that it is the foundation of web 3.0. Quantum computing is not just theoretical now, as can be seen with the development of quantum computers by google (72 qubits), Xanadu (24 qubits), IBM (127 qubits), Intel (49 qubits) etc. Quantum Computers are real threats to blockchain technology, as discussed in the article. Our literature review found that to make blockchain stable even with quantum computers, work must be done at all the layers of blockchain, not just one layer. By that, we can genuinely make a quantum-resistant blockchain.

The focus was mainly on the research questions in the survey, and both the research questions were answered. The security threats on the blockchain are divided based on the layers of the blockchain and based on that we analysed the papers. As shown in Fig 3, most of the work (i.e., 30%) mainly focused on the data layer, which seems likely because mainly encryption and transactions are handled in this layer. The next area of focus was the application and presentation layer, with 24% of articles has shown the work on that. This layer includes applications based on blockchain, which may include smart contracts or chain codes. Therefore, the security of this layer is essential; however, the focus of the articles found concerning specific applications, so the focus should be on general solutions as well. For the network layer, it is found that 23% of papers work to find secure quantum networking and 16% of the articles found work on either changing the consensus algorithm or proposing the new algorithm in itself. Only 1% of articles discussed infrastructure and 6% about working with distributed ledger other than blockchain like IOTA, which is based on the directed acyclic graph.

Next, we aimed to categorise solutions based on four categories only, i.e. Hash-based signatures, Code-based cryptography, lattice-based cryptography and Multivariate cryptography as discussed in [87] for post-quantum cryptography . However, instead of sticking to these four, we decided to make it more transparent and focus on the essential solutions. As shown in Fig. 4, around 25% of papers focused on lattice-based cryptography.

Consensus is necessary for blockchain for the settlement of the transaction. 14% of the papers proposed a new or modified consensus algorithm using either a new hash function, digital signature, or quantum properties. As hash functions and digital signatures are the backbones of blockchain technology, it is necessary to create new or modified signature schemes, and it has been found that 11% of research papers focused on hash-based signatures and 11% of the paper focused on quantum blind signatures. So these are the key areas where research is going on. Analysing the problems and solutions, it is clear that some layers still need some work, like the infrastructure and consensus layers. These layers are also necessary. Findings also suggest that some authors give a solution for one layer and claim that the blockchain will be posted quantum blockchain To make blockchain safe from quantum attacks, it is necessary to create the solution keeping in mind all the layers and find a solution that covers the problems of each layer.

This paper mainly focused on the research found in the literature to increase blockchain security in the post-quantum era. Some literature having reviews based on different focus areas are also found, like authors in [95] focused on proof of stake only, authors in [91] discuss the survival of DLTs after quantum computing. However, it was not thorough, focus on bridging quantum, and classical computing is done in [9], authors in [10] has done a good survey on post-quantum blockchain and compared the significant features of the post-quantum encryption cryptosystems that advanced to the second round of NIST call. This study is restricted for database and year selection to make the review process repeatable and free from any bias.

Even after that, many articles are obtained, evaluated, read, classified, and summarised, and answers to the research questions are presented. The findings suggest that it is needed to see blockchain systems in layers, and researchers should provide solutions to the quantum attacks based on these layers.

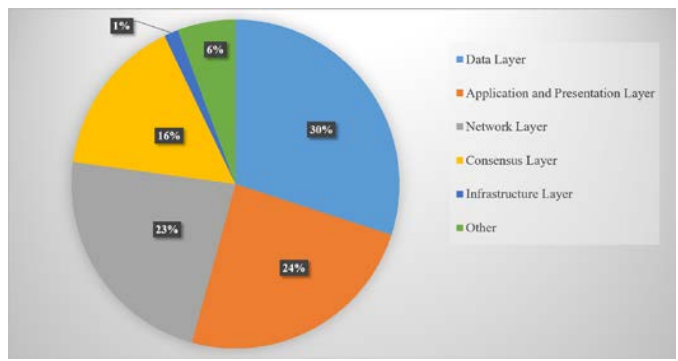

Fig. 3. Security Challenges Identified based on Blockchain Layers.

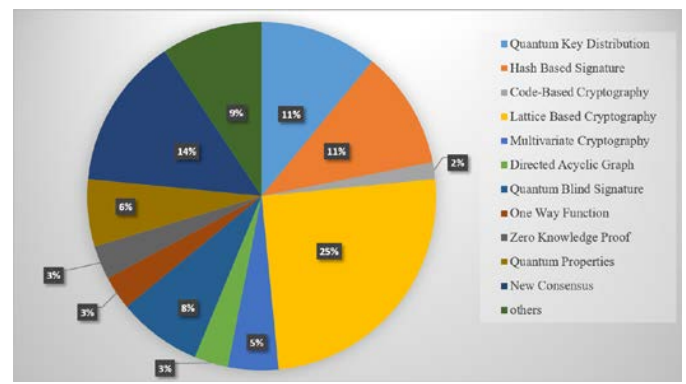

Fig. 4. Frequency Analysis of Solution for the Discussed Challenges.

## V. Conclusion

This study starts with the development of the research questions. To appropriately answer these questions, the systematic literature review is done, and the process is explained in-depth, including the database selection, search process, inclusion and exclusion criterion, creating and extraction of fields and summarising the results. During this process, we found and classified threats based on blockchain layers. Some of the threats were spread over different layers, so these threats are discussed individually for a proper explanation. Many different solutions are also found regarding these threats. The mapping between these threats and solutions has been presented, keeping in mind the full proof solution of post-quantum blockchain.

We discovered that blockchain could operate after quantum computers, but it must work on every layer of the blockchain network, or the solution will not be feasible. Even after developing solutions, they must be thoroughly tested in the real world. If a new application, whether decentralised or not, is being created on the blockchain, quantum attacks should be considered from the planning phase. It has been discovered that blockchain in its current form is unsuitable and must be modified. In the future, researchers will need to create similar solutions and test them for all such issues that have yet to be solved or discussed.